\documentclass{aa} 
\usepackage{graphicx}
\usepackage{txfonts}

\begin{document}

\title{The galaxy counterpart and environment of the dusty damped
Lyman-$\alpha$ absorber at $z=2.226$ towards Q\,1218$+$0832}\titlerunning{The
environment of the dusty DLA towards Q\,1218$+$0832}

\author{
J.~P.~U.~Fynbo\inst{1,2},
L.~Christensen\inst{1,2},
S.~J.~Geier\inst{3,4},
K.~E.~Heintz\inst{1,2},
J.-K.~Krogager\inst{5},
C.~Ledoux\inst{6},
B.~Milvang-Jensen\inst{1,2},
P.~M\o ller\inst{7},
S.~Vejlgaard\inst{1,2},
J.~Viuho\inst{1,2},
G.~{\"O}stlin\inst{8}
}
\institute{
Cosmic DAWN Center,
\email{jfynbo@nbi.ku.dk}
\and
Niels Bohr Institute, University of Copenhagen, Jagtvej 155, 2200 Copenhagen N, Denmark
\and
Instituto de Astrof{\'i}sica de Canarias, V{\'i}a L{\'a}ctea, s/n, 38205, La Laguna, Tenerife, Spain
\and
Gran Telescopio Canaias (GRANTECAN), 38205 San Crist{\'o}bal de La Laguna, Tenerife, Spain
\and
Universit\'e Lyon1, ENS de Lyon, CNRS, Centre de Recherche Astrophysique de Lyon UMR5574, F-69230 Saint-Genis-Laval, France
\and
European Southern Observatory, Alonso de C\'ordova 3107, Vitacura, Casilla 19001, Santiago, Chile
\and
European Southern Observatory, Karl-Schwarzschildstrasse 2, D-85748 Garching, Germany
\and
Department of Astronomy, Oscar Klein Centre, Stockholm University, AlbaNova universitetscentrum, SE-106 91 Stockholm, Sweden
}
\authorrunning{Fynbo et al.}

\date{Received 2023; accepted, 2023}

\abstract{
We report on further observations of the field of the quasar Q\,1218$+$0832.
Geier et al.~2019 presented the discovery of the quasar resulting from a search
for quasars reddened and dimmed by dust in foreground damped
Lyman-$\alpha$ absorbers (DLAs). The DLA is remarkable by having a very large
\ion{H}{i} column density close to 10$^{22}$ cm$^{-2}$. Its dust extinction
curve shows the 2175\AA\  bump known from the Local Group. It also shows
absorption from cold gas exemplified by \ion{C}{i} and CO molecules. For this
paper, we present narrow-band observations of the field of Q\,1218$+$0832 and
also use an archival Hubble Space Telescope (HST) image to search for the galaxy counterpart of the DLA.
No emission from the DLA galaxy is found in either the narrow-band imaging or
in the HST image. In the HST image, we could probe down to an impact parameter
of 0.3 arcsec and a 3-$\sigma$ detection limit of 26.8 mag per arcsec$^2$. In
the narrow-band image, we probed down to a
0 arcsec impact parameter and detected nothing down to a 3-$\sigma$ detection
limit of about 3$\times$10$^{-17}$ erg s$^{-1}$ cm$^{-2}$.  We did detect a
bright Lyman-$\alpha$ emitter 59 arcsec south of Q\,1218$+$0832 with a flux of
3$\times$10$^{-16}$ erg s$^{-1}$ cm$^{-2}$. We conclude that the DLA galaxy
must be located at a very small impact parameter ($<$0.3 arcsec, 2.5 kpc) or it
is optically dark. Also, the DLA galaxy most likely is part of a galaxy group.}    

\keywords{quasars: general -- quasars: absorption lines -- 
quasars: individual: Q\,1218$+$0832}

\maketitle

\section{Introduction}     
\label{sec:introduction}

Considerations about the nature of the first galaxies go back at least to the
1960s. \citet{PP1967} argued that the first galaxies ought to be very bright in
redshifted Lyman-$\alpha$ emission. Some searches were conducted, but except
for the quasars discovered in 1963 \citep{Schmidt1963} and other active
galactic nuclei, no 'normal' galaxies were discovered at redshifts
$z>2$ \citep[e.g.][]{Koo1986}.  In 1986, a breakthrough resulted from the
realisation that normal galaxies could be identified as damped Lyman-$\alpha$
absorbers (DLAs) and studied in absorption against the light of background
quasars \citep{Wolfe1986}. DLAs were detected as broad absorption lines in the
Lyman-$\alpha$ forest with neutral hydrogen column densities in excess of
$2\times10^{20}$ 
cm$^{-2}$ similar to what is measured in the gaseous disks of local spiral
galaxies. The first such object was discovered in a quasar spectrum already in 1972
\citep{Beaver1972}, but at that time their nature as intervening galaxies was not
fully realised. Even today, DLAs remain the most important class of objects for tracing cosmic 
chemical evolution in detail for a large range of chemical elements 
\citep[e.g.][]{Lu1996,Wolfe2005,PH2020}.

The study of $z>2$ galaxies in emission progressed slower than the 
absorption studies. Only very few of the galaxies causing DLAs could be 
detected by their own light \citep[e.g.][and references therein]{MW1993}.
In 1996 the observational study of the first galaxies was revolutionised by the 
discovery, largely driven by the advent of 8-10 m class telescopes, of the so-called 
Lyman break or drop-out galaxies \citep{Steidel1996} and in a sense the current 
plethora of discoveries of galaxies at $z>6$ is a consequence and unfolding of this 
1996 breakthrough.

Now with 50 years having passed since their discovery, there has been great
progress in understanding the nature of DLAs and their relation to galaxies
detected in emission. Around the turn of the millennium, it was realised that
the difficulty in detecting DLA galaxies in emission most likely had to do with
the steepness of the galaxy distribution function. Most of the DLAs must be
caused by dwarf galaxies below the detection limit of galaxy surveys
\citep{Fynbo1999,Haehnelt2000,Schaye2001}. This was somewhat disconcerting as
this meant that galaxies studied in absorption and in emission would remain
almost disjunct samples. Avenues for progress returned when it was realised
that metallicity-luminosity and metallicity-mass correlations most likely were
already established at $z>2$ and were valid for absorption-selected galaxies
too \citep{Moller2004,Ledoux2006,Christensen2014}. Based on the simple model in
\citet{Fynbo2008}, an observing strategy for detecting DLA galaxies in emission
was designed and that led to a substantial increase in the number of detections
\citep{Fynbo2010,Fynbo2013,Noterdaeme2012,Krogager2017}. Subsequently, galaxy
counterparts of metal-rich DLAs have also been detected in emission at
sub-millimetre wavelengths
\citep[e.g.][]{Neeleman2017,Moller2018,Kanekar2020,Kanekar2022}. The use of
Integral Field Units has also resulted in more detections in H-$\alpha$ and
Lyman-$\alpha$ emission \citep{Peroux2011,JW2014,MacKenzie2019,Lofthouse2023}.

The focus on metal-rich DLAs rejuvenated the discussion of dust bias in
connection with DLAs. Since most quasars have been selected based on their
optical colours, it is unavoidable for the most dusty and metal-rich DLAs, due
to dimming and reddening of the background quasars, 
to be under-represented
\citep{Ostriker1984,FallPei1999,WildHewett2005}. \citet{Krogager2019} show,
through detailed simulations of quasar selection and the effect of dust in
DLAs, that the cosmic density of metals may well be underestimated by as much
as a factor of 5 at redshifts around 2.5 due to dust bias. Such a large
uncertainty poses a serious problem when it comes to mapping cosmic chemical
evolution and establishing the connection between DLAs and galaxies observed in
emission. We have therefore worked on various ways to search for these missing
quasars that otherwise have evaded normal quasar selection techniques due to
reddening caused by dust in foreground DLAs
\citep[e.g.][]{Fynbo2013,Krogager2016}. Other teams used radio selection to try
to circumvent the dust bias of optically selected quasar surveys
\citep[e.g.][]{Ellison2001,Jorgenson2006,Sadler2020,Gupta2021}.

In this paper, we present new data for one such system, namely the striking
example of the $z=2.226$ DLA towards the $z=2.60$ Q\,1218$+$0832 \citep[for
other examples see][]{Krogager2016B, Fynbo2017, Heintz2018}.  In
\citet{Geier2019} (paper I hereafter), the discovery of the quasar and the
characterisation of the DLA in terms of \ion{H}{i} column density and other
absorption lines based on low-resolution ($\lambda / \Delta \lambda = 500$ --
$1680$) spectroscopy is described. The DLA has a very large column density
close to 10$^{21.95\pm0.15}$ cm$^{-2}$ and has absorption from both neutral
carbon and CO molecules (due to the low resolution, column densities for these
lines could not be determined). From the low-resolution spectrum, a lower limit
of 10\% solar was placed on the metallicity (based on \ion{Zn}{ii} lines). The
dust extinction curve shows the presence of the 2175-{\AA} bump known from the
Milky Way extinction curve \citep{Stecher1965}. The reddening was determined to
be $A_V = 0.82 \pm 0.02$ mag assuming $z=2.2261$ for the location of the dust.
We have not yet managed to secure a high-resolution spectrum of Q\,1218$+$0832,
which would allow precise measurement of the metallicity and the column
densities of molecules. Instead of pursuing this approach, our paper focusses
on describing new observations of the field using narrow-band imaging. The
primary objective of these observations was to search for the galaxy
counterpart of the DLA and other Lyman-$\alpha$ emitting galaxies in its
environment. Furthermore, we analysed public imaging data obtained with 
the Hubble Space Telescope (HST) that
happen to cover the position of Q\,1218$+$0832. These observations allowed us
to search for broad-band emission from a galaxy counterpart down to small
impact parameters (a few kiloparsecs). 

The paper is organised in the following way. In Sect.~\ref{sec:data}, we
present our observations and briefly describe the data reduction. In
Sect.~\ref{sec:results}, we present our results, which we discuss in the
context of the related results from other teams in Sect.~\ref{sec:discussion}.
In Sect.~\ref{sec:conc}, we present our conclusions.
In the paper we assume a flat $\Lambda$CDM
cosmology with H$_{0}$ = 70 km s$^{-1}$ Mpc$^{-1}$, $\Omega_{\Lambda}=0.7,$ and
$\Omega_{m}=0.3$.

\section{Observations and data reduction}    \label{sec:data}

\begin{figure}[th]
\centering
\includegraphics[scale=0.44]{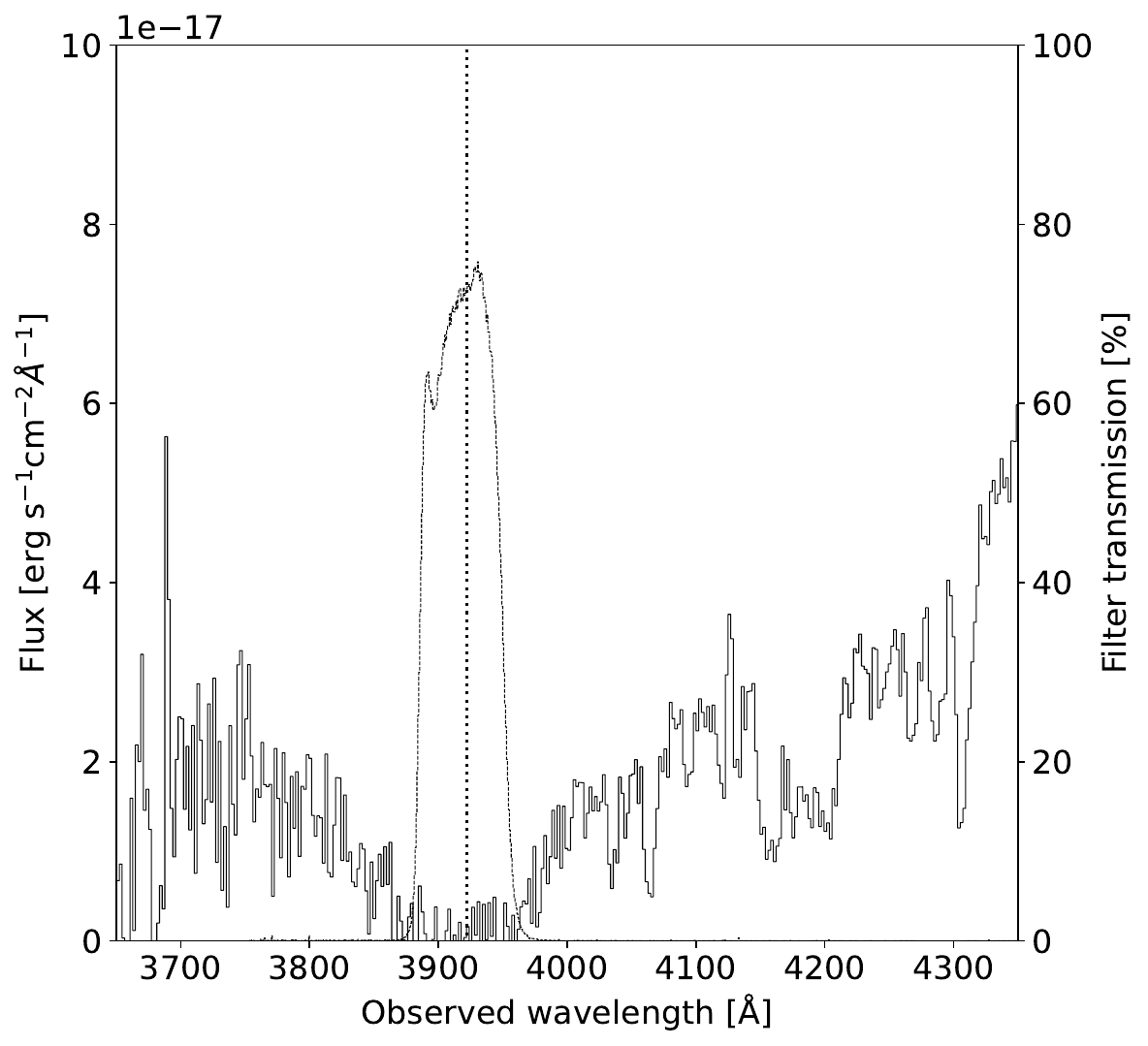}
\caption{
Region around the DLA in the spectrum of
Q\,1218$+$0832 shown together with the filter transmission curve (dotted
line) of the 63{\AA}-wide narrow-band filter NB391\_7 used in this study
(see also
Fig.~2 in \citet{Geier2019}). The vertical dotted line shows the expected position of 
Lyman-$\alpha$ at the redshift of $z=2.2261$ measured from low-ionisation metal lines in 
\citet{Geier2019}. The filter, although designed for other purposes, matches the DLA trough 
perfectly.}
\label{fig:filterfig}
\end{figure}

We have previously used the Nordic Optical Telescope (NOT) for
narrow-band-based searches for galaxies at redshift $z\approx2$
\citep[e.g.][]{Fynbo1999,Fynbo2002}. Such studies take a long time to prepare
as special filters have to be designed and procured for each target.  However,
we realised that a 63{\AA}-wide filter (named NB391\_7) made for other purposes
\citep{Sandberg2015} already exists in the NOT filter set, which by chance is
perfectly suited for a study of the $z=2.226$ DLA towards Q\,1218$+$0832 (see
Fig.~\ref{fig:filterfig}). In March 2023 we carried out a pilot study (in the
sense that we have not reached the detection limit we ultimately would like to
reach) of the field using the SDSS g and r filters (g and r hereafter) and the
NB391\_7 filter in a 3-night observing with the NOT equipped with the
Alhambra Faint Object Spectrograph and Camera (ALFOSC). One night was lost to
bad weather, but we managed to get data on two other nights with clear sky
conditions and seeing between 1 and 1.5 arcsec.

We also secured spectroscopic observations of an emission line source
discovered in the ALFOSC imaging as described in Sect.~\ref{sec:results}. Those
observations were obtained with the 
Optical System for Imaging and low-Intermediate-Resolution Integrated Spectroscopy
(OSIRIS) at the Gran Telescopio Canarias (GTC). We used
the R1000B grism and a 1.23 arcsec-wide slit providing a resolution of $\lambda
/ \Delta \lambda = 500$ and a wavelength coverage of 3600--7770 \AA. The
spectrum was taken at a high airmass of 2.0, as the object was only observable
briefly at the beginning of the night, but otherwise under excellent observing
conditions. 

The log of observations can be seen in Table~\ref{tab:log}. 

\begin{table}[!t]
\centering
\begin{minipage}{0.5\textwidth}
\centering
\caption{Log of ALFOSC and OSIRIS observations.}
\begin{tabular}{lllcc}
\noalign{\smallskip} \hline \hline \noalign{\smallskip}
Observation & Date  & Exposure time \\
       &       &   (sec)      \\
\hline
ALFOSC/NB391\_7  & 20/03/2023 & 5$\times$3000 \\
ALFOSC/NB391\_7  & 22/03/2023 & 2$\times$3000 \\
ALFOSC/g & 20/03/2023 & 5$\times$200  \\
ALFOSC/g & 22/03/2023 & 5$\times$200  \\
ALFOSC/r & 20/03/2023 & 5$\times$200  \\
ALFOSC/r & 22/03/2023 & 5$\times$200  \\
OSIRIS/R1000B & 07/07/2023 & 2$\times$1000 \\
\hline
\noalign{\smallskip} \hline \noalign{\smallskip}
\end{tabular}
\centering
\label{tab:log}
\end{minipage}
\end{table}

The data from the NOT were reduced and combined using standard procedures for
bias subtraction, flat-fielding, and image combination implemented in a set of
Python scripts. The code for image combination using sigma-clipping is
available on GitHub\footnote{\url{https://github.com/jfynbo/Pyclip}}. The
combined seeing in the stacked images are 1.3, 1.2, and 1.2 arcsec for g, r, and
NB391\_7, respectively.

Unfortunately, no standard stars were observed on the nights of 
observation.\ However, based on the calibration described in Sect.~\ref{sec:results},
we estimate that the 3-$\sigma$ detection limit in the narrow-band image
is 3$\times$10$^{-17}$ erg s$^{-1}$ cm$^{-2}$.

The spectroscopic data were reduced using a set of Python scripts 
for the reduction of long-slit spectra. The code is available on GitHub\footnote{\url{https://github.com/keheintz/PyReduc}}.
The spectrum was wavelength-calibrated using HgAr and Ne arc frames. 
The spectrum was flux-calibrated using observations of the spectrophotometric standard star
Ross640 observed on the same night.

We also analysed public HST observations of the field taken with the Advanced
Camera for Surveys (ACS) in the F814W filter as part of the 
Sloan Lens ACS (SLACS) Survey
(SLACS) 
\citep{Bolton2006}. We obtained reduced data from the Hubble Legacy
Archive\footnote{\url{https://hla.stsci.edu}}. The 3-$\sigma$ detection limit in the 
image is 26.8 mag per arcsec$^2$ and the point spread function (PSF) has a full 
width at half maximum (FWHM) of 0.113 arcsec.

We note that three asteroids were detected in our NOT broad-band images: NR10,
OU129, and SA197. All three have well-determined orbits.

\section{Results}    \label{sec:results}

\begin{figure*}[th]
\centering
\includegraphics[scale=0.80]{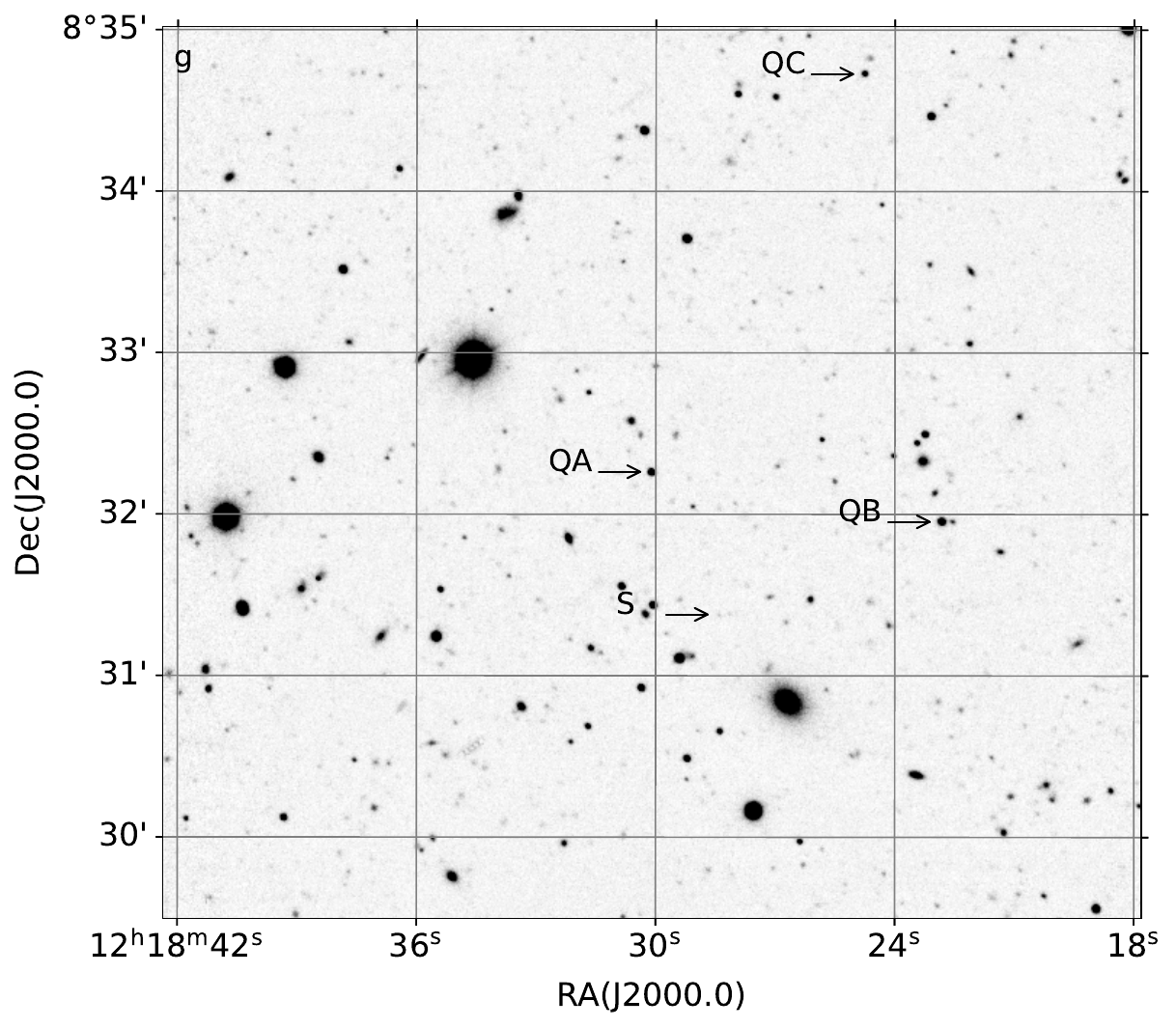}
\caption{Full g-band image around Q\,1218+0832 (marked 'QA').
    The field of view is 6.0$\times$5.4 arcmin$^2$. North is up and east is to
    the left. Other objects marked are discussed further in the text: a
    Lyman-$\alpha$ emitter marked 'S', and two other quasars are marked 'QB'
    and 'QC'.}
\label{fig:field}
\end{figure*}

Fig.~\ref{fig:field} shows the full field covered by our NOT images with the
positions of objects discussed in the paper marked with arrows. 

\subsection{The DLA galaxy counterpart}

\begin{figure}[th]
\centering
\includegraphics[scale=0.50]{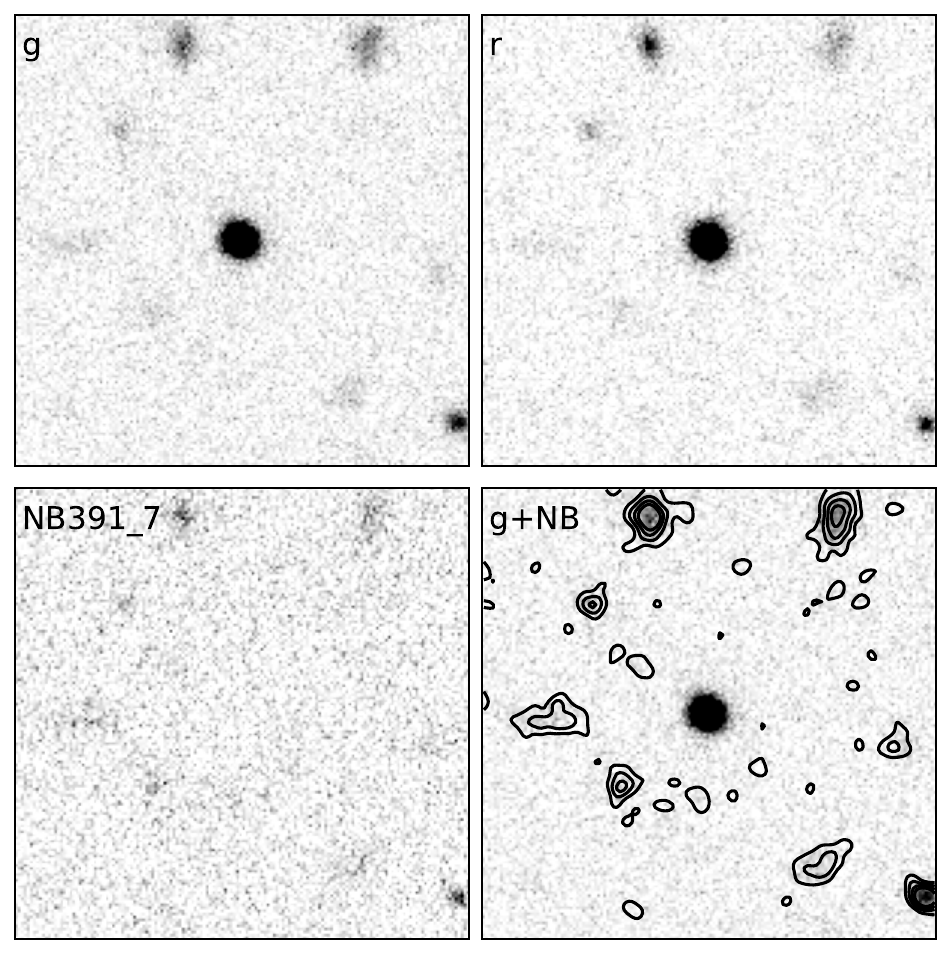}
\caption{ 32$\times$32 arcsec$^2$ region around Q\,1218+0832. In
    the top row, we show the broad-band g and r images. The bottom
    left image shows the NB391\_7 image. In the bottom right image, we overplotted
    the contours of the NB391\_7 image smoothed by a Gaussian kernel with a
    width equal to the seeing on top of the g-band image. We note that the
    quasar is undetected in the NB391\_7 image.}
\label{fig:DLAregion}
\end{figure}

\begin{figure}[th]
\centering
\includegraphics[scale=0.60]{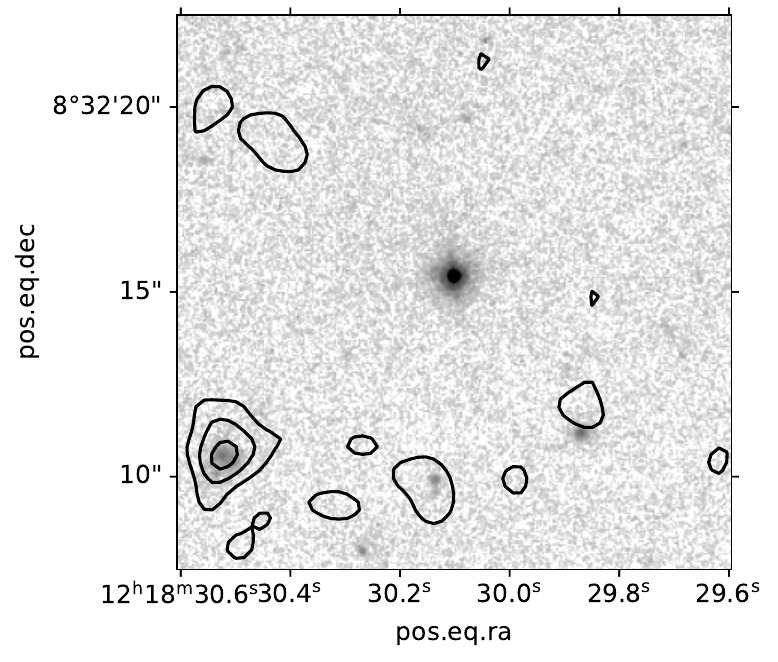}
\caption{DLA region (15$\times$15 arcsec$^2$). The underlying image is the
    HST F814W image. The contours are from the narrow-band image smoothed by a
    Gaussian kernel with width matched to the seeing. }
\label{fig:DLAhstNB}
\end{figure}

In Fig.~\ref{fig:DLAregion} we show a 32$\times$32 arcsec$^2$ around
Q\,1218$+$0832 from our NOT imaging. The quasar is completely absent in the
narrow-band image. A few nearby sources were detected in the narrow-band and
broad-band images. Fig.~\ref{fig:DLAhstNB} shows a 15$\times$15 arcsec$^2$ zoom
in on the region around Q\,1218$+$0832 from the HST F814W image with the
NB391\_7 data overlaid as contours. Some of the weak
contours from the NB391\_7 image south of the quasar overlap with sources
detected in the HST image. We need deeper narrow-band observations to establish
if any of these sources are Lyman-$\alpha$ emitters at $z=2.2261$.

\begin{figure}[th]
\centering
\includegraphics[scale=0.68]{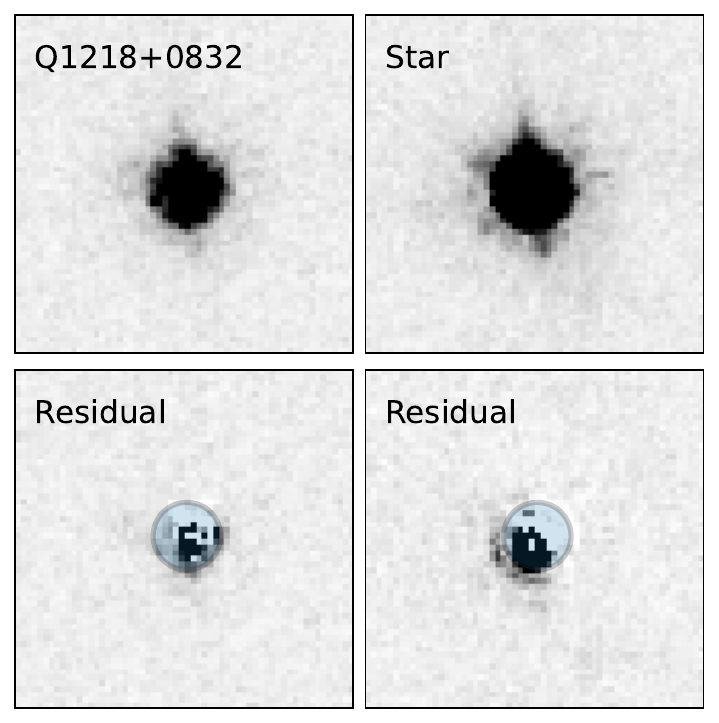}
\caption{3$\times$3 arcsec$^2$ region before (top row) and after (bottom row)
    PSF subtraction. The two left columns show Q\,1218$+$0832 and the two right
    columns show another point source that is about 50\% brighter than
    Q\,1218$+$0832. The is no significant excess emission outside of 0.3 arcsec
    from the quasar (marked by a circle in the lower row).}
\label{fig:psfsub}
\end{figure}

In order to search for continuum emission from the galaxy counterpart of the
DLA at a small impact parameter, we also attempted PSF
subtraction in the HST image. We used GALFIT \citep{Peng2010} to fit and
subtract a PSF defined by an unsaturated star in the field. In
Fig.~\ref{fig:psfsub} we compare the result of the PSF subtraction for
Q\,1218$+$0832 with the results for a nearby star
at RA(J2000.0) = 12:18:25.9, Dec(J2000.0)=$+$08:32:27.0 that is about 50\%
brighter than Q\,1218$+$0832. This comparison allowed us to gauge the size of
the systematic PSF-subtraction residuals. There is no convincing emission
detected down to an impact parameter of about 0.3 arcsec (2.5 proper kpc at
$z=2.2261$). At smaller impact parameters, the PSF subtraction introduces
substantial systematic errors, making it difficult to differentiate between
PSF-subtraction residuals and potential genuine signals originating from the
DLA galaxy and/or the quasar host galaxy. 

\subsection{Lyman-$\alpha$ emitters in the field}

Although we could not derive a precise flux calibration of the NB391\_7 image,
we could still identify objects with a strong excess or a strong deficit of
flux in the narrow band using the colour-colour diagrams as described in, for
example, \citet{MW1993,Fynbo1999}, and \citet{Fynbo2002}. 

\begin{figure}[th]
\centering
\includegraphics[scale=0.54]{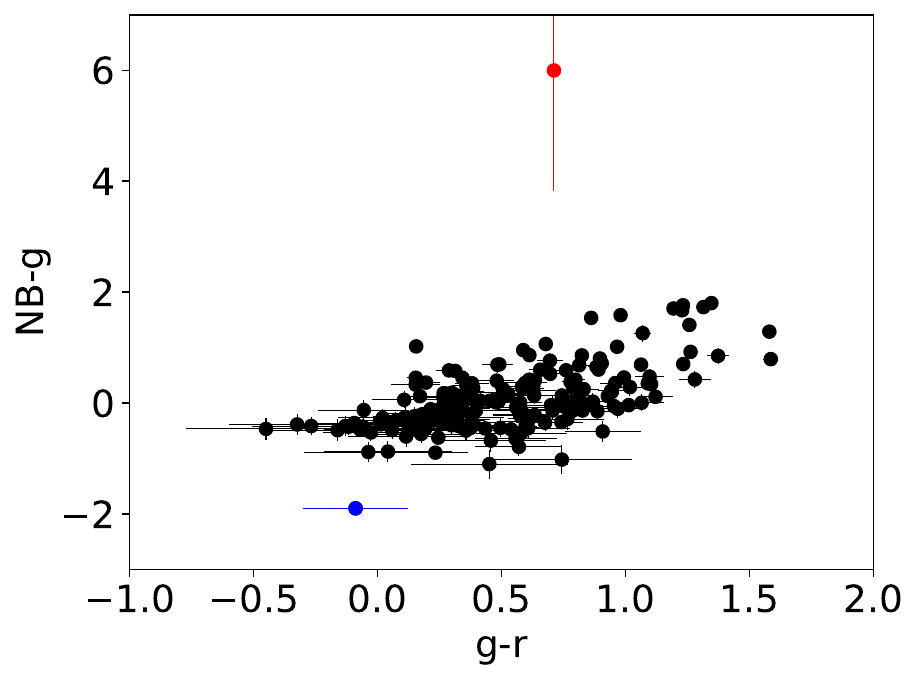}
\caption{Colour-colour diagram based on the NOT imaging in the g and r
    filters and the NB391\_7 filter. Two objects stand out, namely
    Q\,1218$+$0832 (marked in red), which has a deficit of flux in the narrow
    filter, and an emission line source, marked in blue, with excess emission
    in the narrow filter.}
\label{fig:colcol}
\end{figure}

\begin{figure}[th]
\centering
\includegraphics[scale=0.50]{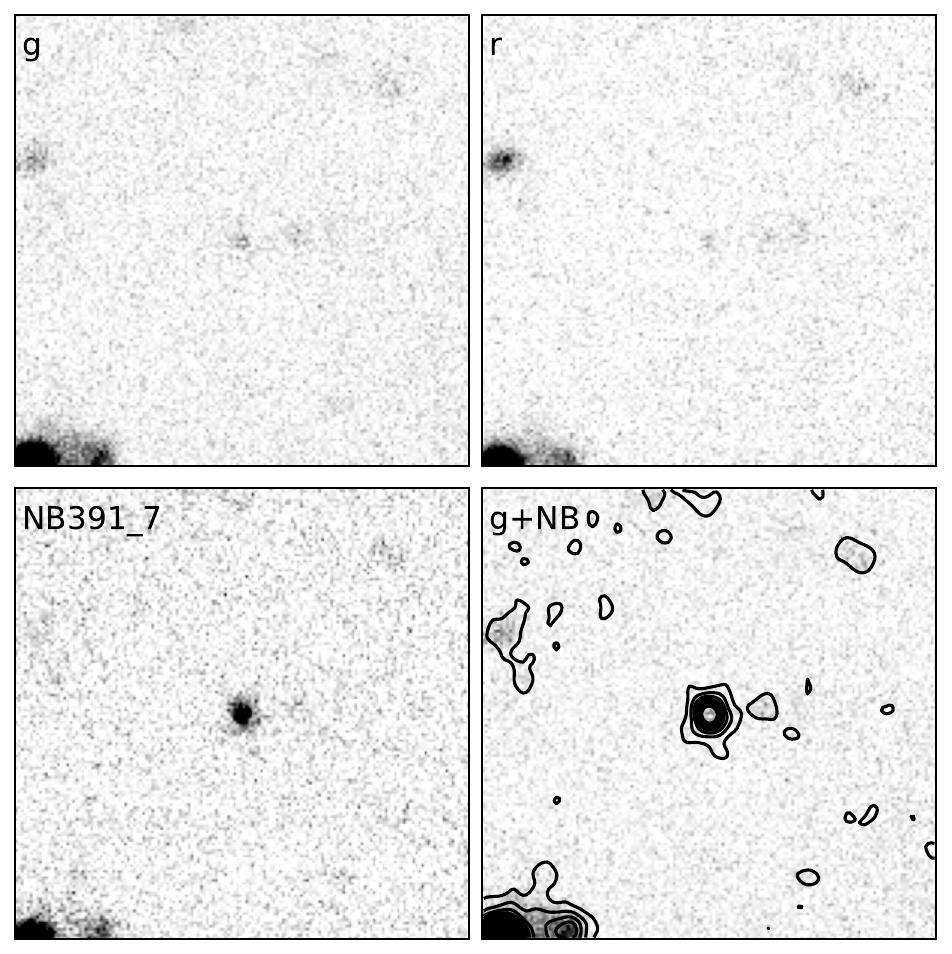}
\caption{32$\times$32 arcsec$^2$ region around the position of
    the strong emission line source we detect 90 arcsec south of Q\,1218+0832.
    In the top row, we show the broad-band g and r images. The bottom left image
    shows the NB391\_7 image. In the bottom right image, we overplotted linearly
    spaced contours of the NB391\_7 image smoothed by a Gaussian kernel with a
    width equal to the seeing on top of the g-band image.}
\label{fig:Emitter}
\end{figure}

Q\,1218$+$0832 stands out as the object with the strongest narrow-band flux
deficit. One object at RA(J2000) = 12:18:28.41,  Dec(J2000)=$+$08:31:22.65 has
a strong excess of emission in the narrow-band image as seen in
Fig.~\ref{fig:Emitter}. The object is marked as 'S' 59 arcsec south of
Q\,1218$+$0832 in Fig.~\ref{fig:field}. At $z=2.23$, 59 arcsec correspond 
to 493 proper kpc. The object is very bright and can be
seen in all the individual 3000-sec narrow-band exposures. No other emission
line sources are detected above 5$\sigma$ significance. 

\begin{figure}[th]
\centering
\includegraphics[scale=0.65]{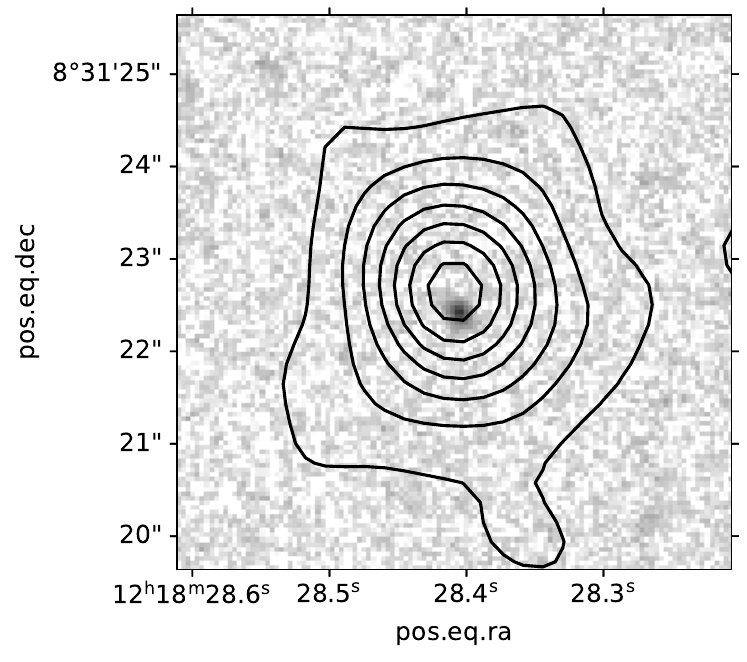}
\caption{ 6$\times$6 arcsec$^2$ region around the emission line
    source. The underlying image is the HST F814W image. The contours are from
    the narrow-band image smoothed by a Gaussian kernel with a width matched to
    the seeing. The contour levels are linearly spaced.}
\label{fig:EmitterhstNB}
\end{figure}

In Fig.~\ref{fig:EmitterhstNB} we overlaid the contours of the narrow-band image
on top of the F814W ACS image. In the HST image, the source is only slightly
larger than the PSF with a measured FWHM of 0.119
arcsec, whereas a PSF has a FWHM of 0.113 arcsec. The intrinsic size of the
object must be only about 0.037 arcsec or 300 pc at $z=2.2261$. In the
narrow band, the source was also resolved, leading to an intrinsic size of about
4 kpc. Hence, the Lyman-$\alpha$ emission is much more extended than the
continuum emission. This is something that is frequently seen for
Lyman-$\alpha$ emitters at these redshifts
\citep[e.g.][]{MW1998,Rauch2008,Steidel2011}.

Assuming that most objects have an NB391\_7 $-$ $g$ colour close to 0, we can
get a rough flux calibration of the NB391\_7 image. Under this assumption, the
emission line source has a magnitude of NB391\_7(AB)$\approx$22.8. The
broad-band magnitudes are $g(AB)$ = 24.38$\pm$0.11 and $r(AB)$ = 24.46$\pm$0.18
measured in a large circular aperture with a 3 arcsec radius. The HST image
gives consistent results for the F814W filter: AB = 24.32$\pm$0.06.  Given the
estimate of the NB391\_7(AB) magnitude, the flux and luminosity for the emission
line source can be estimated using the formulae provided in \citet{Fynbo2002}.
The flux is estimated to be close to 3$\times$10$^{-16}$ erg s$^{-1}$ cm$^{-2}$
and the luminosity close to 1.3$\times$10$^{43}$ erg s$^{-1}$. 

\begin{figure}[th]
\centering
\includegraphics[scale=0.37]{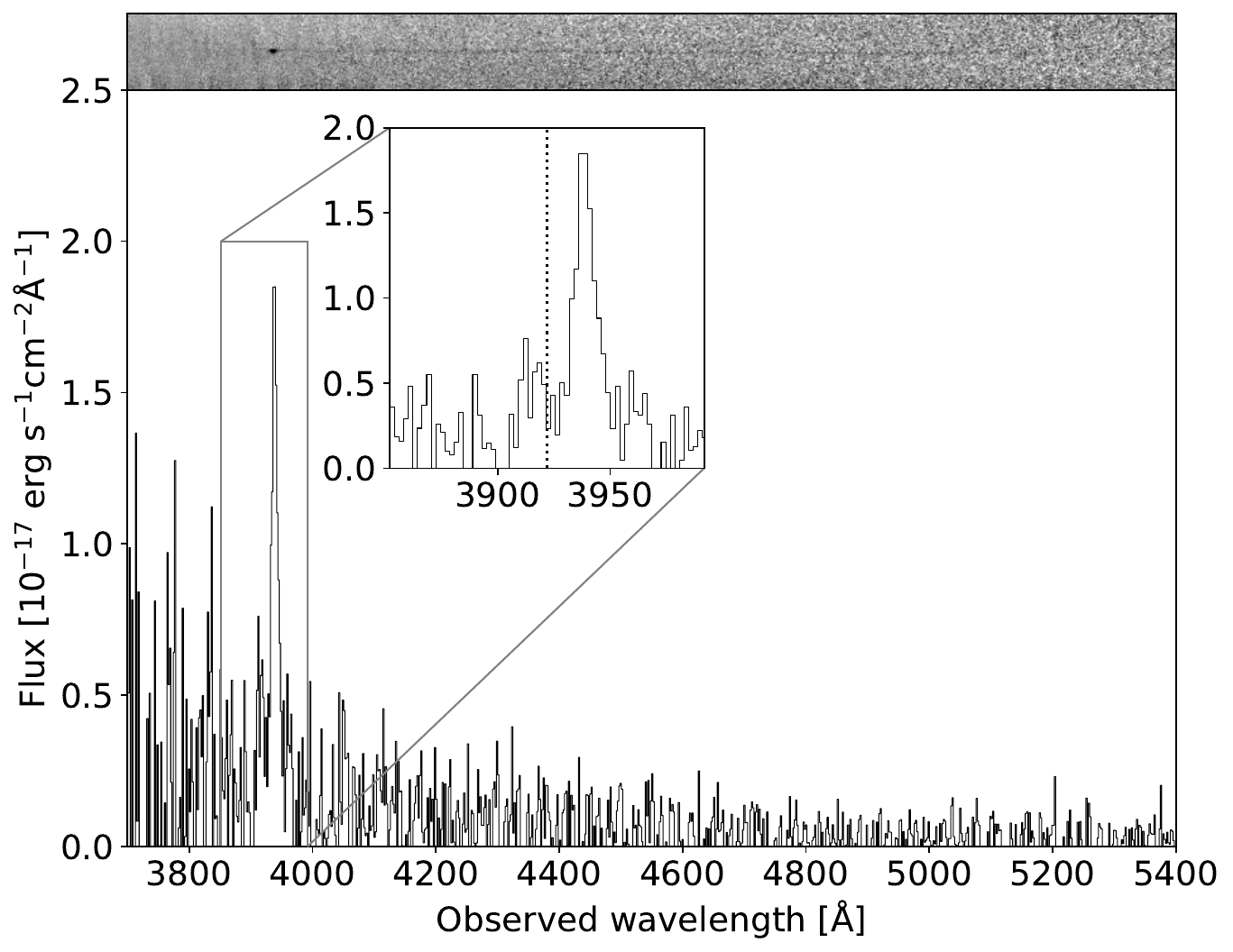}
\caption{Blue end of the OSIRIS/R1000B spectrum of the bright Lyman-$\alpha$ 
marked S in Fig.~\ref{fig:field}. The top panel shows the two-dimensional spectrum and the bottom 
panel the extracted one-dimensional spectrum. A single bright emission line was detected 
at 3937.8$\pm$0.4 \AA \ on top of a blue continuum. The inset shows a zoom-in on the line where
the dotted line shows the position of Lyman-alpha centred at $z=2.2261$.}
\label{fig:OSIRISspec}
\end{figure}

A spectrum was secured in July despite somewhat poor visibility. In 
the spectrum, shown in Fig.~\ref{fig:OSIRISspec}, we detected a single 
bright emission line at 3937.8$\pm$0.4 \AA, which nominally was redshifted by 1218 km s$^{-1}$
relative to Lyman-$\alpha$ at $z=2.2261$. The inset in Fig.~\ref{fig:OSIRISspec} shows 
the profile of the line. Due to resonant scattering, the Lyman-$\alpha$ line is 
often asymmetric and with a double horn profile \citep{Neufeld1990,Pela2009,Verhamme2018}
and the systemic redshift of the galaxy can therefore not be accurately determined from 
the centre of the Lyman-$\alpha$ line, which can be redshifted by up to at least 
1500 km s$^{-1}$ \citep{Shapley2003}. To measure the systemic redshift of the galaxy,
we need measurements of restframe optical emission lines such as Balmer lines or [\ion{O}{iii}]
lines that can be observed in the near-infrared. 
The flux of the line is 2.5$\times$10$^{-16}$ erg s$^{-1}$ cm$^{-2}$, 
which is in good agreement with the estimated flux from the narrow-band imaging 
given the uncertainty in the narrow-band calibration and the effect of slit loss.

\subsection{Other objects of interest in the field}

There is another quasar in the field, namely SDSS\,J121822.83$+$083157.0
(marked 'QB' in Fig.~\ref{fig:field}). This quasar is at nearly the same
redshift as Q\,1218+0832 ($z=2.637$ vs $z=2.60$ for Q\,1218+0832) and has a
separation of 109 arcsec from Q\,1218+0832 and 90 arcsec from the emission line
source. Converted to physical distances at $z=2.23$, the distance between the
two quasar sight lines is 899 proper kpc.

\begin{figure*}[th]
\centering
\includegraphics[scale=0.48]{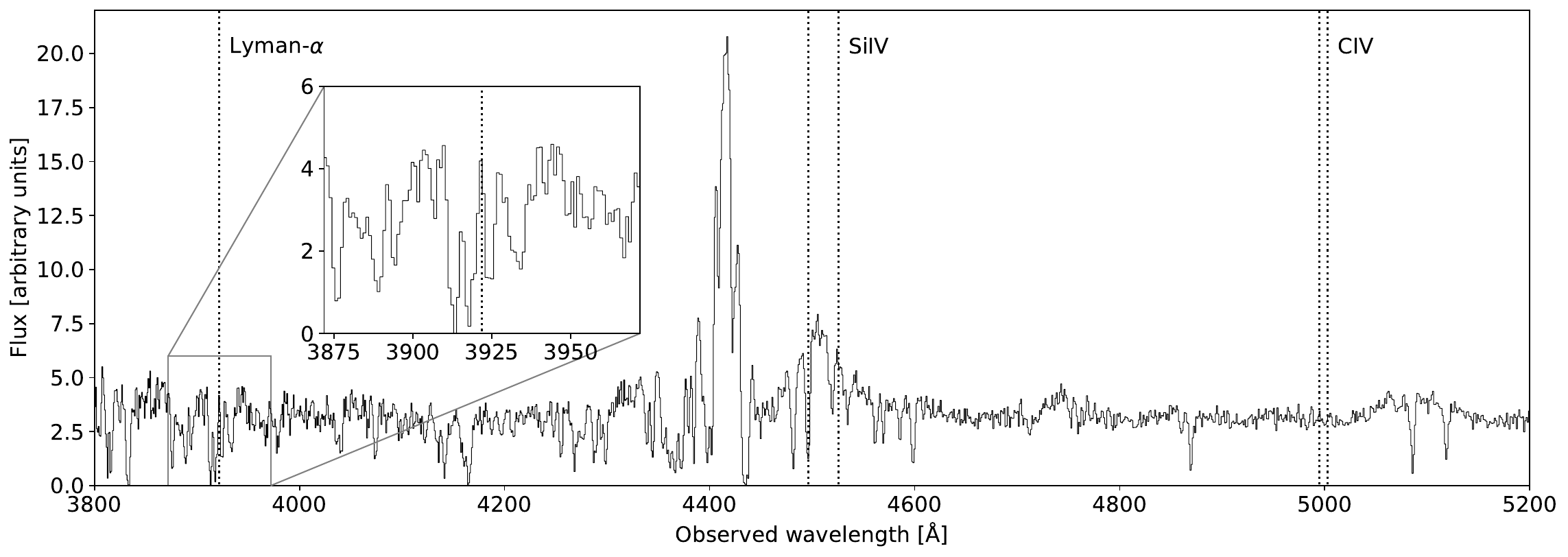}
\caption{Section of the SDSS spectrum of the $z=2.637$ quasar
    SDSS\,J121822.83$+$083157.0 ('QB' in Fig.~\ref{fig:field}) located 109
    arcsec west of Q\,1218$+$0832 ('QA' in Fig.~\ref{fig:field}). Vertical
    lines indicate the expected positions of \ion{H}{i}, \ion{Si}{iv}, and
    \ion{C}{iv} absorption lines at the redshift of the DLA, $z=2.2261$. The
    inset shows a zoom-in on the region around Lyman-$\alpha$. No significant
    absorption was detected at that redshift. The metal lines seen in the
    spectrum are from an associated $z_{abs} > z_{em}$ system at $z=2.650$.}
\label{fig:SDSSquasar}
\end{figure*}

In the spectrum of SDSS\,J121822.83$+$083157.0 (shown in
Fig.~\ref{fig:SDSSquasar}), there is no metal absorption at $z=2.2261$ down to
a 3-$\sigma$ observed equivalent width limit of 0.6 \AA. There
is also no strong absorption in the Lyman-$\alpha$ forest at this redshift.
As shown in the inset in Fig.~\ref{fig:SDSSquasar}, there is Lyman-$\alpha$ absorption just a few hundred km s$^{-1}$ blueshifted
and redshifted relative to  $z=2.2261$, but this is neither damped nor strong
enough to be from a Lyman-limit system (LLS). Rather than absorption, there
seems to be an island of high transmission at $z=2.2261$. 

Furthermore, we note that \citet{Richards2009} classified the object
SDSS\,J121824.76$+$083443.5 as a quasar with a photometric redshift of
$z=2.245$ (marked 'QC' in Fig.~\ref{fig:field}). This object is located 168
arcsec north of Q\,1218$+$0832. The upper and lower values for the photometric
redshift estimate are by \citet{Richards2009} and were determined to be $z_{L} = 2.040$
and $z_{U} = 2.630$, respectively. Hence, the object may well be unrelated to
the DLA, but it would be interesting to measure the spectroscopic redshift for
this source to establish if it is part of the same structure as the DLA or
possibly is at the same redshift as the two other quasars. If it is at a larger
redshift of the DLA, the spectrum could still be used to look for intervening
absorption at $z=2.2261$-- the redshift of the DLA towards Q\,1218$+$0832. We
note that the object has neither been detected by Gaia nor by the Wide-field Infrared 
Survey Explorer (WISE) 
\citep{Wise2015,Gaia2023}.

\section{Discussion} 
\label{sec:discussion}

\subsection{The DLA}

The neutral hydrogen in the structures traced by DLAs is so large, more
specifically similar to the amount of baryons we find in stars in local
galaxies, that DLA must represent an important ingredient in the galaxy
formation recipe \citep{Wolfe1986RSPTA}. However, the precise manner in which
this ingredient should be added to the recipe is still not fully understood.
The absorption characteristics of DLAs as they are imprinted on the light of
the background quasars are very similar to the characteristics of the
absorption seen in gamma-ray burst (GRB) afterglows, which we know result from
absorption in the interstellar medium (ISM) of star-forming galaxies \citep[e.g.][and many similar
cases]{Jensen2001,dUP2018}. The two classes of absorbers also follow the same
scaling relations \citep{Arabsalmani2015}. It is therefore plausible that DLAs
must trace environments that are similar to the ISM of star-forming galaxies.
It is possible to reconcile the absorption statistics of quasar and GRB
absorbers in a simple model where they are drawn from the same underlying
sample of star-forming galaxies at $z>2$ \citep{Fynbo2008,Krogager2020}. DLAs
also allow us to trace important scaling relations
\citep{Moller2013,Christensen2014,Rhodin2018}. Getting similarly detailed
information, for example, about chemical abundances for galaxies detected in emission
would be either impossible or extremely expensive. 

The amount of neutral hydrogen in DLAs and sub-DLAs show hardly any evolution
over a wide redshift range from 5 to 0 \citep{Prochaska2009,
Noterdaeme2009,Noterdaeme2012,Zafar2013,Crighton2015}. Therefore, it is not the
case that there is a one-to-one correspondence between the baryons seen as
DLAs at $z>2$ and the baryons found inside stars in the local Universe. Maybe
DLAs represent a mix of absorption in the ISM phase as well as absorption from
cold gas in circum-galactic material (CGM). The CGM component can itself be a
mix of material accreting onto the dark matter halo of the galaxy and material
leaving the host in the form of a galactic wind. Such a scenario seems to be
supported by simulations of DLAs
\citep{Pontzen2008,Schaye2013,Bird2014,SommerLarsen2017,Rhodin2019}. 

It is important to locate more galaxy counterparts of DLAs as this is the best
way to gain insight into how DLAs in the real Universe concretely are related
to the galaxies we see in emission. Furthermore, for the issue of dust bias, we
would like to know which sight lines are most likely to significantly redden
and obscure background sources as in the case Q\,1218$+$0832 and other similar
cases.  In the last dozen years, good progress has been made in in finding
galaxy counterparts of DLAs, especially when focussing on metal-rich DLAs.
\citet{Krogager2020} represent the most recent compilation of detections. This
compilation, which includes 21 systems, shows that DLA galaxies follow
correlations between the impact parameter of the galaxy and the line of sight
to the background quasar: more metal-rich galaxies tend to have
larger impact parameters and DLAs with larger \ion{H}{i}
column densities tend to have smaller impact parameters.   

The galaxy counterpart of the DLA towards Q\,1218$+$0832 remains
undetected. There
is nothing seen in narrow-band imaging (albeit shallow) and nothing in the HST
image at impact parameters $0.3<b<5$ arcsec (corresponding to about 2.5 -- 40
kpc). Both from the modelling and from the observed sample in
\citet{Krogager2020}, we expect the counterpart of the DLA towards
Q\,1218$+$0832 to be located at a small impact parameter - due to its very large
\ion{H}{i} column density. The presence of \ion{C}{i} and CO absorption also
supports a small impact parameter \citep{KrogagerNoterdaeme2020}, but we do
note that cold gas was also detected in the $z=2.583$ DLA towards
Q\,0918$+$1636 that is located at an impact parameter of 2.0 arcsec (16 kpc).
The galaxy is not a strong Lyman-$\alpha$ emitter; otherwise, it would have
shown up both in the long-slit spectroscopy presented in paper I and in the
narrow-band imaging presented here. The most promising outlook for detecting
the galaxy is either through near-infrared spectroscopy, which would allow the
detection of H$\alpha$ or [\ion{O}{iii}] emission as for the DLAs towards
Q\,0918$+$1636 \citep[][neither of which are bright in Lyman-$\alpha$
emission]{Fynbo2013}, or with the Atacama Large Millimeter/submillimeter Array (ALMA)
where the galaxy could show up in
CO emission as in the case of J2225$+$0527 \citep{Kanekar2020}, which is
detected in CO(3-2) emission at an impact parameter of 5.6 kpc. 

It is interesting that the quasar is completely absent in the narrow-band
imaging. We can use this to draw conclusions about the transverse size of the
DLA. If there really is no emission detectable with the NB391\_7 filter, the
DLA must cover not only the quasar broad line region but also the quasar host
galaxy (assuming that the host has emission at 1090 {\AA}, which is the
wavelength probed by the NB391\_7 filter at the redshift of Q\,1218$+$0832). Of
course, we need deeper imaging to infer if this is indeed the case.

Systems such as Q\,1218$+$0832 and its intervening DLA show that the dust
reddening of quasars and other distant sources due to dust in the foreground
objects discussed by \citet{Ostriker1984} is a real effect. We note that the
upcoming Purely Astrometric Quasar Survey \citep[PAQS][]{Krogager2023} will be
able to determine quantitatively how strong the effect is, that is, what the
frequency is for significant reddening and dimming of background quasars due to
dust in foreground galaxies.

\subsection{The environment}

Even though we have not been able to locate the DLA galaxy, we have gained new
insights into its environment. We detected a single bright emission line source
59 arcsec (corresponding to 493 proper kpc at $z=2.2261$) south of
Q\,1218$+$0832. The source is among the brightest Lyman-$\alpha$ emitters at
similar redshifts \citep[compare, e.g. with the samples in
][]{Fynbo2002,Nilsson2009,Sandberg2015,Matthee2021}. It would be interesting to
secure a significantly deeper image in the NB391\_7 filter to reach the
flux level of more typical Lyman-$\alpha$ emitters and hence be able to infer
more about the galaxy density in the field. Comparing to the sample of \citet{Nilsson2009}
(in particular their fig.~3), it is clear that source S is at the absolute
bright end of the luminosity function for Lyman-$\alpha$ and that more typical
galaxies (in that study) are about 2 magnitudes fainter. 
Previous studies of Lyman-$\alpha$
emitters in the fields of high-column density \ion{H}{i} absorbers have found an
indication of filamentary structures or galaxy overdensities
\citep[e.g.][]{MW1998,Moller2001,Fynbo2003,Lofthouse2023}. In the case of Q\,0918$+$1636, a
bright CO emitter was detected at a distance of 117 kpc from the DLA galaxy
\citep{Fynbo2018}. If DLA galaxies are frequently located in dense galaxy
fields, this could help explain the surprisingly large correlation length of
DLAs \citep{correlation2018}. A preference for DLAs in group environments could
result from tidal stripping of neutral gas out of galaxies in such environments
\citep[See also][]{Bouche2012,Bouche2013,Rauch2016}.

The environment can also be probed by the multiple sight lines offered by
at least two quasars in the field (the quasar nature of
SDSS\,J121824.76$+$083443.5 remains to be established by spectroscopy). There
is no metal-line absorption at $z=2.2261$ in the spectrum of the $z=2.637$
quasar SDSS\,J121822.83$+$083157.0. In the Lyman-$\alpha$ forest, there is no
absorption exactly at $z=2.2261$, but some forest lines are slightly blueshifted and
redshifted relative to that redshift.  \citet{DOdorico2002} studied ten similar
cases of multiple nearby quasar sight lines and detected five out of ten
matching systems of strong absorption systems within 1000 km s$^{-1}$,
indicating an overdensity of strong absorption systems over separation lengths
from~ 1 to 8 h$^{-1}$ Mpc. More recently, \citet{Stawinski2023} have studied a
larger sample of 32 DLAs intersecting close pairs of quasars. Their sample
probes smaller impact parameters out to about 300 kpc. At these distances, there
is a covering factor of $>90$\% for \ion{C}{iv} and $>50$\% for strong
\ion{H}{i} absorbers (LLSs and DLAs). Similar studies have been carried out
testing the covering factor of \ion{C}{iv} absorption around Lyman-break
galaxies and Lyman-$\alpha$ emitters \citep{Adelberger2005, Bielby2017,
Muzahid2021, Dutta2021, Banerjee2023,Galbiati2023}. These studies have found excess \ion{C}{iv}
absorption at distances of a few hundred proper kiloparsecs.     

\section{Conclusions}
\label{sec:conc}

The objective of this paper is to present new information about the dusty
$z=2.2261$ DLA towards the reddened quasar Q\,1218$+$0832. This information
constitutes of two main parts. Concerning the galaxy counterpart of the DLA, we
have analysed new narrow-band observations and archival HST imaging to place
strong constraints on its properties. It is most likely located at a very low
impact parameter of $b<0.3$ arcsec (or $<2.5$ proper kpc at $z=2.2261$).
Concerning the environment of the DLA galaxy, we have shown that there is at
least one other galaxy nearby, namely a bright Lyman-$\alpha$ emitter that is
located 59 arcsec towards the south at a projected distance of 493 proper kpc
at $z=2.2261$. The environment can be further probed using sigh lines to
another quasar in the field 109 arcsec towards the west (899 proper kpc at
$z=2.2261$) and here there is no metal or strong \ion{H}{i} absorption at the
the redshift of the DLA. A third quasar in the field has not been observed
spectroscopically, but the photometric redshift evidence suggests that it could
either be at the same redshift as the DLA or at a higher redshift and hence it can
be used to further probe the environment of the DLA. 

Further spectroscopic observations are required to search for the DLA galaxy at
small impact parameters, to measure the precise redshift of the Lyman-$\alpha$
emitter, and to establish the relation of the third quasar to the environment
of the DLA.

\begin{acknowledgements}
    We thank an anonymous referee for a very helpful report. 
    JPUF thanks Claudio Grillo for his advice about the HST observations.  Based on
    observations made with the Gran Telescopio Canarias (GTC) and with the 
    Nordic Optical Telescope (NOT), installed at 
    the Spanish Observatorio del Roque de los Muchachos belonging to the 
    Instituto de Astrofísica de Canarias. Also based on observations made with the NASA/ESA
    Hubble Space Telescope, and obtained from the Hubble Legacy Archive, which
    is a collaboration between the Space Telescope Science Institute
    (STScI/NASA), the Space Telescope European Coordinating Facility
    (ST-ECF/ESA) and the Canadian Astronomy Data Centre (CADC/NRC/CSA).
    Funding for the Sloan Digital Sky Survey V has been provided by the Alfred
    P. Sloan Foundation, the Heising-Simons Foundation, the National Science
    Foundation, and the Participating Institutions. SDSS acknowledges support
    and resources from the Center for High-Performance Computing at the
    University of Utah. The SDSS web site is \url{www.sdss.org}.  SDSS is
    managed by the Astrophysical Research Consortium for the Participating
    Institutions of the SDSS Collaboration, including the Carnegie Institution
    for Science, Chilean National Time Allocation Committee (CNTAC) ratified
    researchers, the Gotham Participation Group, Harvard University, Heidelberg
    University, The Johns Hopkins University, L’Ecole polytechnique
    f\'{e}d\'{e}rale de Lausanne (EPFL), Leibniz-Institut f{\"u}r Astrophysik
    Potsdam (AIP), Max-Planck-Institut f{\"u}r Astronomie (MPIA Heidelberg),
    Max-Planck-Institut f{\"u}r Extraterrestrische Physik (MPE), Nanjing
    University, National Astronomical Observatories of China (NAOC), New Mexico
    State University, The Ohio State University, Pennsylvania State University,
    Smithsonian Astrophysical Observatory, Space Telescope Science Institute
    (STScI), the Stellar Astrophysics Participation Group, Universidad Nacional
    Aut\'{o}noma de M\'{e}xico, University of Arizona, University of Colorado
    Boulder, University of Illinois at Urbana-Champaign, University of Toronto,
    University of Utah, University of Virginia, Yale University, and Yunnan
    University.  The Cosmic Dawn Center (DAWN) is funded by the Danish National
    Research Foundation under grant No. 140.  JPUF is supported by the
    Independent Research Fund Denmark (DFF--4090-00079) and thanks the
    Carlsberg Foundation for support.  LC is supported by the Independent
    Research Fund Denmark (DFF-2032-00071).  KEH acknowledges support from the
    Carlsberg Foundation Reintegration Fellowship Grant CF21-0103.
\end{acknowledgements}

\bibliographystyle{aa}

\object{GQ\,1218$+$0832, SDSS\,J121822.83$+$083157.0, SDSS\,J121824.76$+$083443.5, NR10, 
OU129, SA197}
\end{document}